\documentclass[journal]{IEEEtran}
\usepackage{url}
\usepackage[utf8]{inputenc}
\usepackage{xcolor}
\usepackage{amsmath}
\usepackage{amssymb}
\usepackage{microtype}

\usepackage{graphicx}
\usepackage{tablefootnote}
\usepackage{booktabs}
\usepackage{tabularx}
\usepackage{tikz}
\usepackage{pgfplots}
\pgfplotsset{compat=newest} 
\pgfplotsset{plot coordinates/math parser=false} 
\newlength\fheight
\newlength\fwidth
\usetikzlibrary{plotmarks,patterns,decorations.pathreplacing,backgrounds,calc,arrows,arrows.meta,spy,matrix}
\usepgfplotslibrary{patchplots,groupplots}
\usepackage{tikzscale}

\usepackage{multirow}
\usepackage{tkz-kiviat}
\usepackage[htt]{hyphenat}

\usepackage[font=footnotesize]{subcaption}
\usepackage[font=footnotesize]{caption}

\usepackage{mathtools}

\usepackage{dblfloatfix}    
\usepackage{colortbl}

\usepackage{makecell}
\usepackage{diagbox}
\usepackage{tikz-qtree}
\usetikzlibrary{trees} 

\usepackage{cite}
\usepackage{bbold}
\usepackage{bbm}

\usepackage{array}
\newcolumntype{?}{!{\vrule width 1.5pt}}
\usepackage{makecell}
\usepackage{mdframed}
\usepackage[many]{tcolorbox}
\usepackage{empheq}

\newtcbox{\mybox}[1][]{nobeforeafter,math upper,tcbox raise base,
  enhanced,frame hidden,boxrule=0pt,interior style={top color=green!10!white,
  bottom color=green!10!white,middle color=green!50!yellow},
  fuzzy halo=1pt with green,drop large lifted shadow,#1}

\usepackage{siunitx}
\usepackage[acronyms,nonumberlist,nopostdot,nomain,nogroupskip]{glossaries}

\newacronym{vanet}{VANET}{Vehicular Ad Hoc Network}
\newacronym{ukf}{UKF}{Unscented Kalman Filter}
\newacronym{cdf}{CDF}{Cumulative Distribution Function}
\newacronym{iid}{IID}{Independent and Identically Distributed}
\newacronym{ctra}{CTRA}{Constant Turn Rate and Acceleration}
\newacronym{kf}{KF}{Kalman Filter}
\newacronym{pf}{PF}{Particle Filter}
\newacronym{hmm}{HMM}{Hidden Markov Model}
\newacronym{its}{ITS}{Intelligent Transport System}
\newacronym{ml}{ML}{Machine Learning}
\newacronym{svm}{SVM}{Support Vector Machine}
\newacronym{nn}{NN}{Neural Network}
\newacronym{gps}{GPS}{Global Positioning System}
\newacronym{dr}{DR}{Dead Reckoning}
\newacronym{bf}{BF}{Bayesian Filtering}
\newacronym{rbf}{RBF}{Radial Basis Function}
\newacronym{c-its}{C-ITS}{Connected and Intelligent Transportation System}
\newacronym{sumo}{SUMO}{Simulation of Urban MObility}
\newacronym{dsrc}{DSRC}{Dedicated Short Range Communication}
\newacronym{csmaca}{CSMA/CA}{Carrier Sense Multiple Access with Collision Avoidance}
\newacronym{qoi}{QoI}{Quality of Information}
\newacronym{aoi}{AoI}{Age of Information}
\newacronym{mac}{MAC}{Medium Access Control}
\newacronym{pdf}{PDF}{Probability Density Function}

\begin{document}
\title{Quality-Aware Broadcasting Strategies for \\ Position Estimation in VANETs}

\author{\IEEEauthorblockN{Federico Mason,\hspace{-0.0mm}$^*$ Marco Giordani,\hspace{-0.0mm}$^*$ Federico Chiariotti,\hspace{-0.0mm}$^*$ Andrea Zanella,\hspace{-0.0mm}$^*$\\Takamasa Higuchi,\hspace{-0.0mm}$^\dagger$ Onur Altintas,\hspace{-0.0mm}$^\dagger$ Michele Zorzi$^*$} \\\vspace{0.22cm}
\IEEEauthorblockA{$^*$Department of Information Engineering, University of Padova -- Via Gradenigo, 6/b, 35131 Padova, Italy\\$^\dagger$Toyota InfoTechnology Center, Inc. -- 94043 Mountain View,
CA, USA\\Email: {\tt\footnotesize\{masonfed, giordani, chiariot, zanella, zorzi\}@dei.unipd.it,\{ta-higuchi, onur\}@us.toyota-itc.com}
\vspace{-0.5cm}}}

\author{\IEEEauthorblockN{Federico Mason, Marco Giordani, Federico Chiariotti, Andrea Zanella, Michele Zorzi} \\
\IEEEauthorblockA{Department of Information Engineering, University of Padova -- Via Gradenigo, 6/b, 35131 Padova, Italy
\\Email: {\tt\{masonfed, giordani, chiariot, zanella, zorzi\}@dei.unipd.it
\vspace{-0.5cm}}}}

\maketitle

\begin{abstract}
The dissemination of vehicle position data all over the network is a fundamental task in Vehicular Ad Hoc Network (VANET) operations, as applications often need to know the position of other vehicles over a large area. 
In such cases, inter-vehicular communications should be exploited to satisfy application requirements, although congestion control mechanisms are required to minimize the packet collision probability. In this work, we face the issue of achieving accurate vehicle position estimation and prediction in a VANET scenario.
State of the art solutions to the problem try to broadcast the positioning information periodically, so that vehicles can ensure that the information their neighbors have about them is never older than the inter-transmission period. However, the rate of decay of the information is not deterministic in complex urban scenarios: the movements and maneuvers of vehicles can often be erratic and unpredictable, making old positioning information inaccurate or downright misleading. 
To address this problem, we propose to use the Quality of Information (QoI) as the decision factor for broadcasting. We implement a threshold-based strategy to distribute position information whenever the positioning error passes a reference value, thereby shifting the objective of the network to limiting the actual positioning error and guaranteeing quality across the VANET.
The threshold-based strategy can reduce the network load by avoiding the transmission of redundant messages, as well as improving the overall positioning accuracy by more than 20\% in realistic urban~scenarios.
\end{abstract}


\section{Introduction}\label{sec:intro}
In recent years, there has been a growing interest in \glspl{vanet}, which have rapidly emerged as a means to enable safer traveling and improved traffic management, and to support infotainment applications~\cite{hartenstein2008tutorial}. As vehicular technologies evolve towards the support of more safety-critical applications, research efforts have been made towards the design of novel \gls{vanet}  architectures and implementations which guarantee timely and accurate positioning of vehicles,\footnote{Although positioning is typically provided by on-board \gls{gps} receivers, \emph{data fusion}  techniques have also been considered in \glspl{vanet} by combining several positioning strategies (including, but not limited to, dead reckoning, map matching, and camera image processing) into a single solution that is more robust and precise than any individual approach~\cite{Balico:2018}.} a fundamental prerequisite  for safety systems as well as for Internet access and multimedia~services.

The  unique characteristics of \glspl{vanet} might cause rapid dynamics and unpredictable changes in the network topology, thereby requiring regular position updates to be disseminated as timely as possible, i.e., ideally at the very same instant they are generated.
However, next-generation \glspl{c-its} and the heterogeneity of their requirements constrain the amount of information that can be successfully broadcast over bandwidth-constrained communication channels. High connectivity  pressure on the \gls{vanet} would likely lead to packet collisions and potentially degrade the accuracy and timeliness of \gls{vanet} services~\cite{joo2018wireless}.

In this scenario, the traditional strategy is to have each vehicle broadcast periodic updates~\cite{kaul2012real} with its positioning information. 
Other vehicles in its communication range will then have the guarantee that the time from the last update never passes the inter-transmission period.
However, the unpredictable variability of the VANET topology might make a periodic broadcasting strategy inefficient in terms of the absolute positioning error, thereby calling for innovative and more sophisticated information distribution~solutions that explicitly consider the \gls{qoi}~\cite{bisdikian2013quality} instead of using the time between subsequent updates as a proxy for~it.
Generally, congestion-avoidance mechanisms have also been proposed in the literature to regulate information distribution  as a function of the network load. 
However, these techniques dynamically adapt the \gls{vanet} transmission parameters, e.g., by controlling the number of neighboring vehicles~\cite{caizzone2005power} or by assigning  different priorities to vehicles with different operating conditions~\cite{huang2009analysis}, regardless of the level of  positioning accuracy that is achieved from the information that is successfully delivered.  
The concept of \emph{value-anticipating} vehicular networking has also been investigated as a means to  efficiently disseminate data in resource-constrained vehicular networks, i.e., by   discriminating the importance of the different positioning information sources, in order to use the limited transmission resources in a way that maximizes the utility for the target applications\cite{higuchi2019value}. 
The value assessment process should be computationally efficient, so that it can be completed in low latency even with the limited on-board computational resources of mid-range and budget car models.

Following this rationale, in this paper we face the challenge of ensuring accurate vehicle position estimation and prediction while minimizing the network load in a cost-effective way.
In this regard, among the original contributions of this paper, we
\begin{itemize}
	\item design a \emph{threshold-based} broadcasting algorithm which \emph{(i)} estimates the positioning error of the vehicle and of its neighbors within communication range, based on purely predictive \gls{ukf} tracking operations, and \emph{(ii)} forces vehicles to distribute state information messages in case the estimated error of the previously broadcast positioning information is above a predefined threshold. The performance of the proposed strategy is compared with that of an elementary \emph{constant inter-transmission period} policy, with which vehicles broadcast state information updates at regular intervals.
	\item investigate the impact of the channel conditions on the overall position estimation accuracy. 
	Inter-vehicular broadcasting operations are modeled based on a realistic implementation of the IEEE 802.11p frame structure.
	\item  investigate the performance of the proposed broadcasting scheme as a function of the \gls{vanet} dynamics. In order to have realistic movement of the vehicles, we generate mobility traces using \gls{sumo}~\cite{krajzewicz2012recent}, an open road traffic simulator designed to handle and model the traffic of large road networks.
	\item evaluate the performance of the \gls{ctra} motion model, which was traditionally proven to be one of the most appropriate models to track vehicular mobility, considering realistic urban \gls{vanet}~scenarios.
\end{itemize}

Our results show that the proposed \emph{threshold-based} broadcasting algorithm, in spite of its simplicity, can improve the position estimation accuracy by more than 20\% compared to state of the art approaches. 
We also demonstrate the impact of the \gls{vanet} topology dynamics on the overall network performance, and  prove that information can decay at different rates depending on how well movements fit the broadcast model, making the time since the last update an imperfect measure of the real \gls{qoi}.
Moreover, we illustrate how the \gls{ctra} model might not represent a suitable solution to characterize long-term vehicular mobility, especially in dynamic urban environments, in spite of its demonstrated success at short-term tracking. The success of the threshold-based strategy is due to its ability to compensate for these limitations, and our theoretical analysis supports this~explanation.

\section{Related work}\label{sec:related}

Estimating the position of  \gls{vanet} nodes can be considered an extension of the target tracking problem~\cite{Boukerche2008vehicular}. In a \gls{vanet}, each vehicle makes use of on-board sensors and wireless communication to collect data regarding the surrounding environment. 
The gathered data can be used as the input of a tracking system that has, as a target state, the set of positions of all the \gls{vanet} nodes. 
Since inter-vehicle communication allows vehicles to share information, the performance of the described tracking system becomes highly dependent on the cooperation between \gls{vanet} nodes. 
An example of a tracking strategy in VANETs is given in \cite{Ramos:2012}, and most research works dedicated to vehicle position estimation and prediction are based on similar paradigms. The specificity of each implementation mainly depends on the motion model chosen to represent the behavior of the  vehicles and on the \gls{bf} algorithm used to process the input data. An analysis of the main motion model used in vehicle tracking is given in \cite{Schubert:2008}. The best-known \gls{bf} algoritms used for the same purpose are the \gls{kf} \cite{Kalman:1960} and the \gls{pf} \cite{Moral:1996}. A vehicle tracking framework based on the \gls{ukf} algorithm \cite{Wan:2000} and the \gls{ctra} motion model is presented in\cite{Lytrivis:2011}. Route information and digital map data are processed by a \gls{pf} algorithm in \cite{Peker:2011}. In \cite{Akabane:2017}, vehicle position forecasting is achieved by exploiting a system based on a \gls{hmm} \cite{Poritz:1988} and the Viterbi algorithm \cite{Viterbi:1967}. 
We highlight that the performance of all the \gls{bf}-based systems strongly depends on the filter settings, which must be defined \emph{a priori}. A comprehensive analysis of the possible \gls{kf} configurations for vehicle tracking is given in \cite{Roth:2014}. 

Conventional tracking approaches mainly focus on the real-time estimation of the target state. 
However, most advanced \gls{c-its} applications also require the prediction of the future motion of the target vehicle. Long-term forecasting can be achieved by simply applying the predictive step of a \gls{bf} filter to the last available state estimation, although this solution does not provide good performance when the behavior of the vehicle is not properly represented by the chosen motion model. 
To overcome this issue, more sophisticated approaches have been proposed in the literature. In \cite{Kang:2017}, the output of a \gls{kf} system is used to achieve the parametric interpolation of target vehicles' future paths. In \cite{King:2018}, the \gls{dr} technique is implemented in order to improve packet forwarding in a highway scenario. Another possibility consists in describing vehicle position prediction as a time series forecasting problem \cite{Balico:2018}. 
\gls{ml} techniques, such as \glspl{svm} \cite{Sapankevych:2009} and \glspl{nn} \cite{Katarya:2018}, can then be designed to improve state estimation.
In \cite{Wang:2015}, \glspl{svm} are used to forecast vehicle trajectories during the time period in which the \gls{gps} signal is not available. In \cite{Park:2011} a \gls{nn} system trained with historical traffic data is used to predict vehicles' future speeds. 
Although \gls{ml} approaches generally guarantee accurate tracking, they require a large amount of sensory observations for the training and suffer from significant computational complexity.

\gls{bf} and \gls{ml} approaches are often combined with the aim of maximizing the performance of the vehicle position estimation and prediction. 
A trajectory prediction system that makes use of an \gls{hmm} module to determine which maneuver a vehicle is performing and an \gls{svm} module to achieve the position prediction is presented in \cite{Deo:2018}. In \cite{Hermes:2009}, a \gls{rbf} classifier is used to compute the setting of a \gls{pf}, which is then exploited to achieve a long-term motion prediction. In \cite{Houenou:2013}, the results of a maneuver recognition system and a tracking system based on the \gls{ctra} motion model are combined together. 

In general, the performance of tracking systems strongly depends on the accuracy of the input data and on how often new data are available. In this perspective, our work tries to minimize broadcasting operations while ensuring accurate position estimation.

\section{System model and broadcasting strategies}\label{sec:model}

In this section we present our system model.
In particular, Sec.~\ref{sub:general_model} describes the \gls{vanet} from an analytical point of view, Sec.~\ref{sub:error_function} presents the error function which is used as the accuracy metric for the performance evaluation, Secs.~\ref{sub:tracking_system} and \ref{sub:channel_model} describe the tracking system implemented by the vehicles and the vehicular network channel model, respectively, and  Sec.~\ref{sub:broadcasting_strategies}  presents the broadcasting strategies we propose. 

\subsection{General model}
\label{sub:general_model}

We represent a \gls{vanet} as a Euclidean graph  $G=(V, E, r)$, i.e., an undirected graph whose vertices are points on a Euclidean plane~\cite{Balico:2018}. $V$ represents the set of nodes, $E$ represents the set of edges and $r$ is the node's communication range.
We say that two vehicles $v_i\in V$ and $v_j\in V$, $i\neq j$, are connected by connection $<v_i,v_j>$ if the distance $d_{i,j}$ between them  is shorter than the communication range $r$, i.e.,  $E=\{<v_i,v_j>:i\neq j\wedge d_{i,j}<r\}$. 
Since the composition of the edge set depends on the position of the vehicles, the topology of the \gls{vanet} is time-variant, e.g., new edges can be activated or disabled according to vehicles movements. 

By representing the network as a Euclidean graph, we assume that the vehicles are moving in a two-dimensional space; while not always realistic, this hypothesis does not compromise the accuracy of our analysis. Moreover, we assume that time can be divided into discrete timeslots. 
To highlight the time dependency of the \gls{vanet}, we denote by $G(t)=(V(t),E(t),r)$ the network graph at time $t$. It becomes intuitive to define the neighbor set $N_i(t)$ of vehicle $v_i$ at time $t$ as the set of vehicles connected to $v_i$ by an edge in $E(t)$: $N_i(t)=\{v_j\in V(t):<v_i,v_j>\in E(t)\}$.

\begin{figure}[t]
  \centering
  \setlength{\belowcaptionskip}{-5pt}
  \includegraphics[height=1.8in]{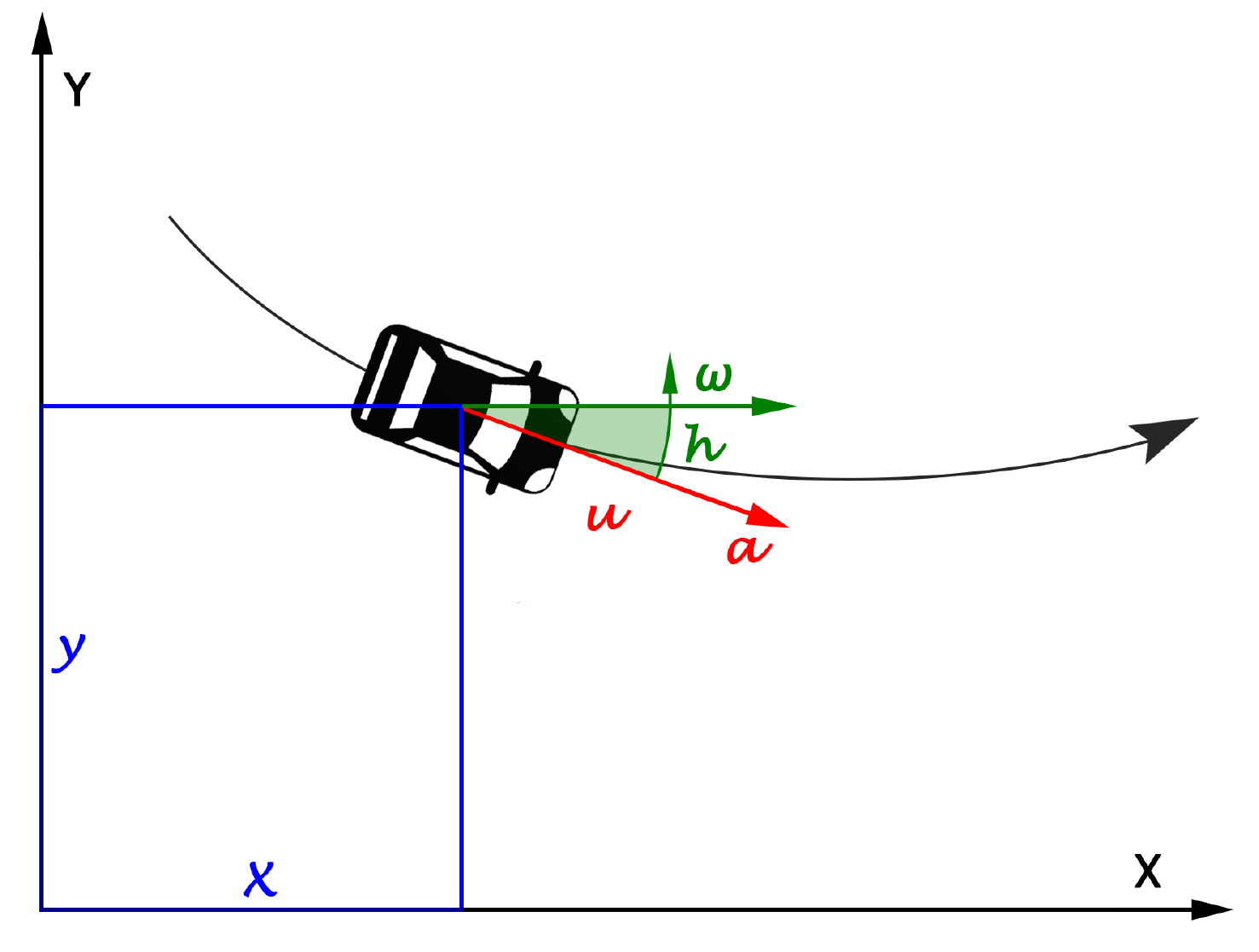}
  \caption{Graphical representation of the  vehicle state $s(t)=\Big(x(t),y(t),h(t),u(t),a(t),\omega(t)\Big)$ at time $t$.}
  \label{fig:vehicle_state}
\end{figure}

The behavior of each vehicle $v_i$ in the VANET at time $t$ is represented by a 6-tuple $s(t)=\Big(x(t),y(t),h(t),u(t),a(t),\omega(t)\Big)$, which we call \emph{vehicle state}. In particular, $x$ and $y$ are the Cartesian coordinates of the vehicle on the road topology, $h$ is the vehicle's heading direction, $u$ and $a$ are the vehicle's tangent velocity and acceleration, respectively, and $\omega$ is the vehicle's  angular velocity. A graphical representation of the vehicle state is given in Fig.~\ref{fig:vehicle_state}. 
Since our work focuses on position estimation and prediction, we define the distance between two states $s_i(t)$ and $s_j(t)$  as the physical distance between the positions of vehicles $v_i$ and $v_j$, $i\neq j$, at time $t$, i.e., $d(s_1(t),s_2(t)) = \sqrt{(x_1(t) - x_2(t))^2 + (y_1(t) - y_2(t))^2}$. 

\subsection{Error function}
\label{sub:error_function}

Consider a reference vehicle $v_i$ $\in$ $V(t)$, called the \emph{ego vehicle}, which is connected to each vehicle $v_j\in$ $N_i(t)$, at time $t$. 
We call $\hat{N}_i(t)$ the subset of $N_i(t)$ containing the vehicles which are tracked by the ego vehicle, i.e., the target vehicles, and $\hat{s}_{i,j}(t)$ the state estimate of $v_j$ carried out by $v_i$ at time $t$.  
Under these hypotheses, the performance, in terms of position estimation accuracy, of the ego vehicle can be assessed by an \emph{error function} $\mathcal{F}(v_i,t)$, i.e., 
\begin{equation} \label{eq:error_function}
\begin{split}
\mathcal{F}(v_i,t) = & \frac{1}{ |\hat{N}_i (t)|+1 } \Big( \lambda_{i,i}(t) d(\hat{s}_{i,i} (t), s_i (t)) + \\ 
& \sum_{ v_j \in  \hat{N}_i (t) } \lambda_{i,j}(t) d(\hat{s}_{i,j} (t), s_j (t)) \Big).
\end{split}
\end{equation}

In \eqref{eq:error_function}, $d(\hat{s}_{i,i} (t), s_i (t))$ represents the error made by $v_i$ in estimating its own state $s_i(t)$, $d(\hat{s}_{i,j} (t), s_j (t))$ represents the error made by $v_i$ in estimating the neighbor state $s_j(t)$, $|\hat{N}_i (t)|+1$ represents the total number of estimations carried out by $v_i$ and $\lambda_{i,j}(t)$ is the logistic function defined as

\begin{equation} \label{eq:sig_factor}
\lambda_{i,j}(t) = A + \frac{K-A} {\left( C + Q e^{ -B (d(s_i(t), s_j (t))-d_0)} \right) ^{1 / \nu }}.
\end{equation}
 Parameters in \eqref{eq:sig_factor}  characterize the logistic function's shape and guarantee that errors referred to spatially close vehicles, the~most safety-constrained neighbors, are weighted more than others. Their values will be detailed in Sec.~\ref{sec:analysis}.
To evaluate the performance of the whole VANET, we define $\mathcal{F}(V, t)$ as the average of $\mathcal{F}(v_i,t)$ among all vehicles~$v_i \in V(t)$:

\begin{equation}\label{eq:err_function_vanet}
\mathcal{F}(V, t) = \frac{1}{|V(t)|}\sum_{v_i \in V(t)}\mathcal{F}(v_i,t).
\end{equation} 

\subsection{Tracking system}
\label{sub:tracking_system}

To minimize the positioning error defined in \eqref{eq:error_function}, at every timeslot the ego vehicle must estimate its state and the state of every other vehicle in the set $\hat{N}_{i}(t)$. 
To reach this goal, the ego vehicle exploits both the information gathered by its on-board sensors and the information received from its neighbors through inter-vehicle communications. To allow the estimation of $s_i(t)$, we assume that, at every timeslot, the ego vehicle's on-board sensors provide a new observation $o(t)$ of $s_i(t)$. Hence, the ego vehicle can model the evolution of its own state through a Bayesian approach, obtaining the system

\begin{equation} \label{eq:bayesian_model}
\begin{cases}
s(t+1) = f(s(t)) + \mu(t), \\
o(t) = m(s(t)) + \nu (t).
\end{cases}
\end{equation}

In \eqref{eq:bayesian_model}, the first equation describes the evolution of the vehicle state $s(t)$ over time, while the second equation describes the relation between $s(t)$ and the state observation $o(t)$. In particular, $f$ is a function describing the \gls{ctra} motion model given in \cite{Tsogas:2005}, while $m$ is a function representing the vehicle's measurement system operation. Moreover, $\mu(t)$ and $\nu (t)$ represent the process and measurement noises, respectively, and are modeled as Gaussian processes with zero mean and covariance matrices $Q$ and $R$. 
We highlight that $Q$ is an identity matrix multiplied by a constant, i.e., $Q= q \cdot I$, while $R$ is a diagonal matrix whose values correspond to the accuracy of the vehicle's measurement system. 
Once all the parameters  in \eqref{eq:bayesian_model} are defined, the ego vehicle can estimate its own state by using a \gls{bf} algorithm. In our model, each vehicle implements a \gls{ukf} algorithm exploiting the \textit{sigma points} parameterization given in~\cite{Julier:2002}. 
The \gls{ukf} is a widely used \gls{bf} method that allows to estimate the state of a system that evolves according to a non-linear model.
By exploiting the \gls{ukf} and the system equations given in \eqref{eq:bayesian_model}, the ego vehicle can  obtain the state estimate $\hat{s}_{i,i}(t)$  of its own state $s(t)$ at each timeslot $t$. 
The \gls{ukf} framework alternates two working steps. During the predictive step the state estimate is propagated through the \gls{ctra} equations; during the updating step the state estimate is updated with the new information provided by the observation~$o(t)$.

To allow $v_i$ to estimate the states of the other \gls{vanet} nodes, each vehicle $v_j$ $\in$ $V(t)$ broadcasts its last estimated state $\hat{s}_{j,j}(t)$ and the related covariance matrix. The time frame by which new transmissions are initiated  depends on the selected broadcasting strategy, as described in Sec.~\ref{sub:broadcasting_strategies}. Each message transmitted by $v_j$ is received by all the vehicles in $N_j(t)$ after a certain communication delay. 
If, by this process, a new neighbor is detected, the ego vehicle initializes a new \gls{ukf}  having as initial state and uncertainty the received state and covariance matrix, respectively. 
The new filter propagates the initial state over time by exploiting only its \textit{predictive step}. In case the ego vehicle receives a packet from a vehicle $v_j$ who was already detected, the \gls{ukf} assigned to $v_j$ is updated with the new data contained in the packet. 

\subsection{Channel model}
\label{sub:channel_model}

Inter-vehicle communications are modeled following the IEEE 802.11p standard, which supports the Physical (PHY) and \gls{mac} layers of the  \gls{dsrc} transmission protocol~\cite{li2010overview}. 
\gls{dsrc} defines seven different channels at the PHY layer, each constituted by $N_{sc,tot}=52$ sub-carriers \cite{Jiang:2008}. For simplicity of discussion, we assume that only a limited number of sub-carriers $N_{sc}<N_{sc,tot}$ can be used for broadcasting state information messages, while the rest of the sub-carriers is used by other VANET applications. 

\gls{dsrc} implements the \gls{csmaca} scheme at the \gls{mac} layer, where nodes listen to the wireless channel before sending. \gls{csmaca} allows for reduced signaling overhead with respect to other IEEE 802.11 standards and enables uncoordinated channel access. We consider a 1-persistent system: when a vehicle senses that its chosen channel is occupied, it attempts to send the positioning update in the next timeslot.
On the other hand,  contention-based access is prone to the \emph{hidden node} problem, which may result in packet collision if an out-of-range vehicle is transmitting towards the same potential receiver.
In our model, we assume that if, in any timeslot, a vehicle $v_i$ receives two packets sent by two other vehicles using the same sub-carrier during the same timeslot,  both packets are discarded.

\subsection{Broadcasting strategies}
\label{sub:broadcasting_strategies}

In our model two different broadcasting strategies for the distribution of the vehicles' state information are implemented.

\paragraph{Constant inter-transmission period policy}
 
A constant inter-transmission period $T_p$ is chosen, so that vehicles initiate new transmissions at regular time frames. This approach is already implemented by most \gls{vanet} positioning applications and  represents the benchmark solution of our~analysis.

\paragraph{Threshold-based policy}

In this case, vehicles try to adjust the broadcasting operations according to the specific characteristics of the \gls{vanet} in which they are deployed. On a practical level, new transmissions are initiated only when the expected position estimation error of the neighboring vehicles overcomes a predetermined threshold $E_{thr}$. To reach this goal, the ego vehicle defines a \gls{ukf} algorithm, as described in Sec.~\ref{sub:tracking_system}, which replicates the working behavior of all  \gls{vanet} nodes that are tracking the ego vehicle itself. 
In this case, the ego vehicle's state is propagated by the  \gls{ukf} algorithm over time by using only its \textit{predictive step}, and is updated every time the ego vehicle carries out a new transmission. 
Hence, in any timeslot the ego vehicle has available both the \textit{a posteriori} state estimate $\hat{s}_{i,i}(t)$, which is the output of its main filter, and the \textit{a priori} state estimate $\hat{s}_{i,i}^p(t)$, which is the output of its purely predictive filter.  In each timeslot, the ego vehicle compares $\hat{s}_{i,i}(t)$ and $\hat{s}_{i,i}^p(t)$: if the state difference $d(\hat{s}_{i,i}(t), \hat{s}_{i,i}^p(t))$ is bigger than $E_{thr}$, a new transmission is immediately initiated. 
The threshold $E_{thr}$ is a function of the traffic density and is selected through an exhaustive approach. In particular, we consider $N_{E_{thr}}$ different values of the error threshold and select the one that minimizes the positioning error.


We highlight that the behavior of both  policies can change according to some specific events. First, both  strategies define a maximum inter-transmission period, i.e., a maximum number of consecutive timeslots during which no transmissions are initiated. Second, both  strategies force the ego vehicle to broadcast immediately its position information every time it receives a message from an undetected vehicle. This aims at reducing the number of vehicles that are undetected, i.e., the vehicles $v_j$ which belong to $N_i(t)$ but not to $\hat{N}_i(t)$. 

\section{Performance analysis}\label{sec:analysis}

\begin{table}[t!]
\begin{tabular}{@{}lll@{}}
\toprule
Parameter            & Value                 & Description                                                \\ \midrule
$T_{\rm sim}$            & $300$ s                 & Simulation duration                                        \\
$T_t$                & $100$ ms                 & Timeslot duration                                          \\
$r$                  & $140$ m                 & Communication range                                        \\
$T_{d}$             & $100$ ms                 & Communication delay
 \\
$d_0$                & $42$ m                  & Safety distance                                          \\
$v_{max}$            & $13.89$ m/s             & Max. speed in VANET                                     \\
$d_v$                & $120$ vehicles/km$^2$   & Vehicular density                                          \\
$A_S$                  & $0.5168$ km$^2$         & VANET area size                                            \\
$E_{thr}$ & $\{0,\dots, 42\}$ m & Error threshold \\
$T_{p}$ & $\{0,\dots, 10\}$ s & Inter-transmission period \\
$|V|$                & $62$                    & Number of VANET nodes                                     \\
\{A, K, C, Q, B, $\nu\}$ & \{$1$, $0$, $1$, $1$, $0.05$, $0.2$\} & Logistic function params                                         \\
 \bottomrule
\end{tabular}
\caption{Simulation parameters.}
\label{tab:params}
\end{table}

\begin{table}[t!]
\centering
\setlength{\belowcaptionskip}{-0.5cm}
\begin{tabular}{@{}lll@{}}
\toprule
Parameter & Value & Description \\ \midrule
$R_{1,1}$ & $1.18535$ m$^2$ & Position accuracy along x \\
$R_{2,2}$ & $1.18535$ m$^2$ & Position accuracy along y \\
$R_{3,3}$ & $0.5$ (m/s)$^2$ & Speed accuracy \\
$R_{4,4}$ & $0.39$ (m/s$^2$)$^2$ & Acceleration accuracy \\
$R_{5,5}$ & $0.09211$ rad$^2$ & Heading accuracy \\
$R_{6,6}$ & $0.01587$ (rad/s) $^2$ & Turn rate accuracy \\ \bottomrule
\end{tabular}
\caption{Accuracy for vehicle state parameters.}
\label{tab:accuracy}
\end{table}

The simulation parameters have been chosen based on~realistic system design\hspace{0.10cm}considerations\hspace{0.10cm}and are\hspace{0.10cm}summarized in\hspace{0.09cm}Table~\ref{tab:params}.
\paragraph{General parameters} 
\label{par:general_parameters}
We use conservative IEEE 802.11p PHY and MAC layer parameters for our simulations, which yield  a maximum discoverable range of $r=140$ m~\cite{benin2012vehicular}, while the communication delay is set to $T_d =100$ ms, corresponding to one timeslot $T_t$.
Our results are derived through a Monte Carlo approach, where multiple independent simulations of duration $T_{\rm sim}=300$ s are repeated to obtain different statistical quantities of interest.

\paragraph{Position estimation parameters} 
\label{par:ukf_parameters}
In Sec.~\ref{sec:model} we assessed that the behavior of each vehicle in the VANET can be fully represented by its  state $s(t)$, whose parameters are affected by a non-negligible measurement noise which is modeled as a Gaussian process with zero mean and covariance matrix $R$. The elements of $R$ are reported in  Table~\ref{tab:accuracy} and are derived from the models in  \cite{Kim:2014,Driver:2007,Falco:2017}.
Moreover, we assign to the logistic function parameters in~\eqref{eq:sig_factor} the following values: $A=1$ and $K=0$, $C=1$, $Q=1$, $B=0.05$, $\nu=0.2$, and $d_0=42$ m (in this way, $\lambda_{i,j}(t) \simeq 1$ as $d(s_i(t), s_j (t)) \rightarrow 0 $ and $\lambda_{i,j}(t) \rightarrow 0$ as $d(s_i(t), s_j (t)) \rightarrow + \infty $).
In particular, $d_0$ represents the \emph{safety distance} that must be held in an urban scenario, and determines the threshold beyond which  $\lambda_{i,j}(t)$ starts decreasing.



\begin{figure}[b!]
\centering
 \begin{subfigure}[t]{0.44\columnwidth}
  \centering
  \includegraphics[width=0.99\columnwidth]{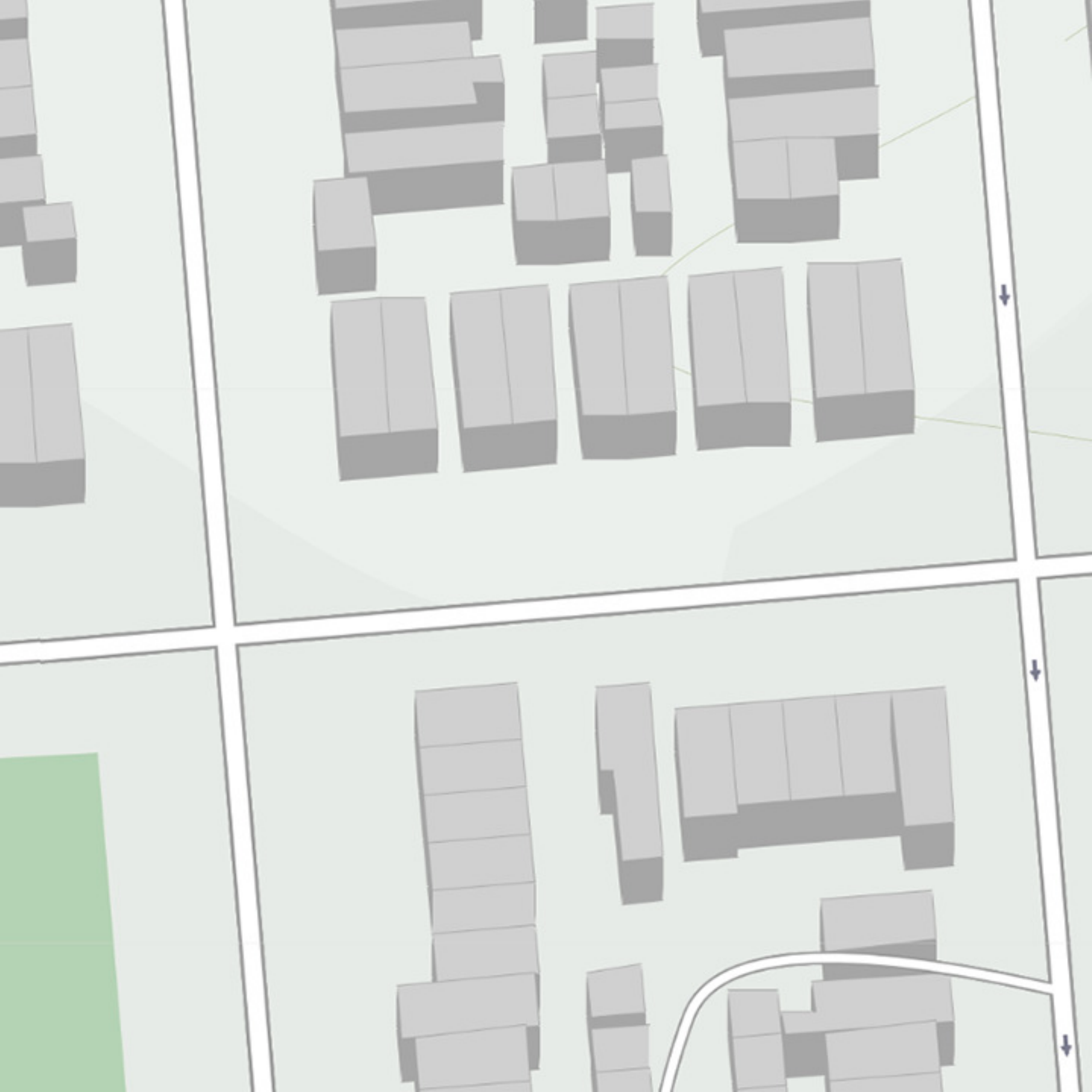}
    \caption{Openstreetmap scenario.}
  \label{fig:osm}
  \end{subfigure} 
  \begin{subfigure}[t]{0.44\columnwidth}
  \centering
  \includegraphics[width=0.99\columnwidth]{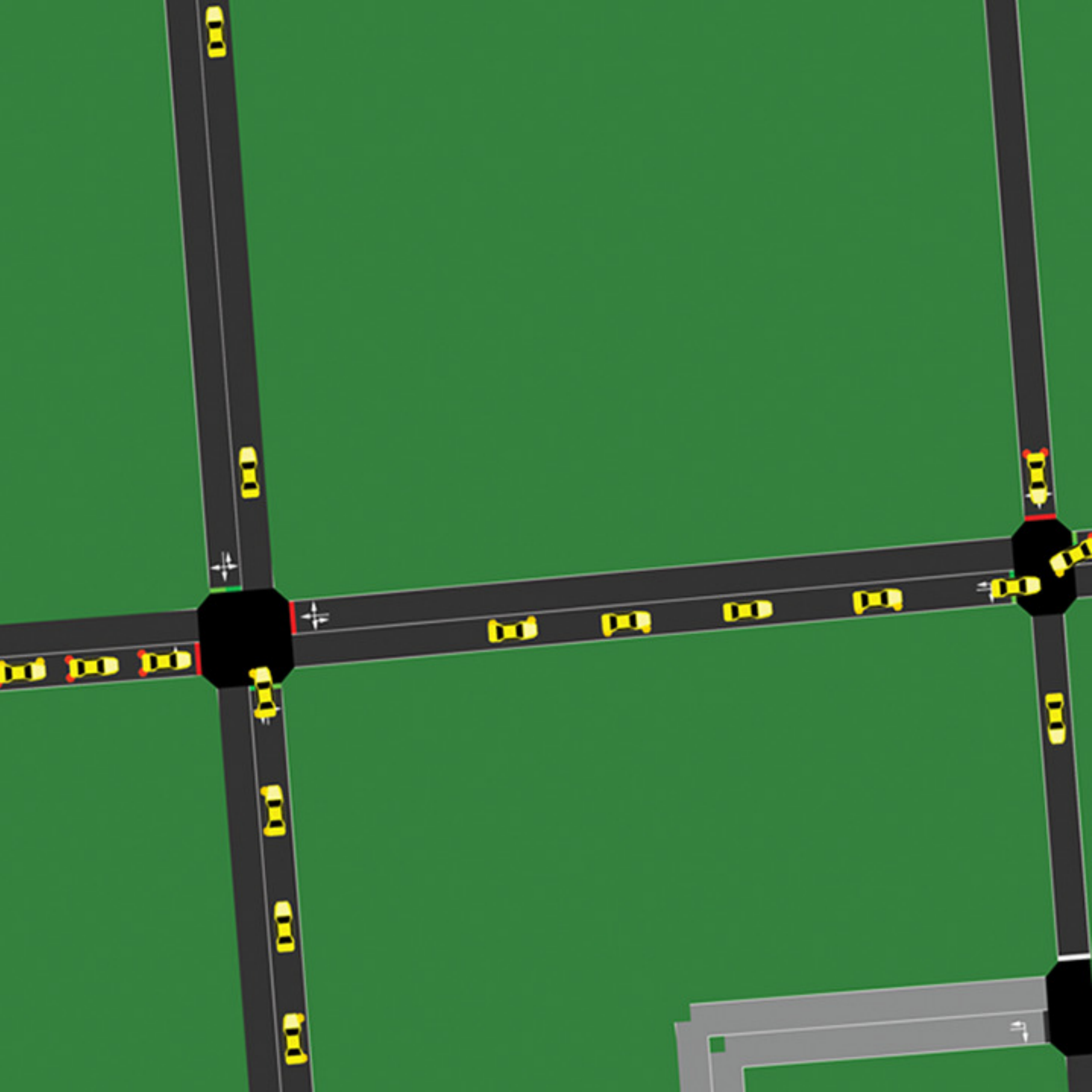}
  \caption{SUMO scenario.}
    \label{fig:sumo}
  \end{subfigure}
  \caption{Representation of a portion of the urban map that is considered for the performance evaluation.}
  \label{fig:map}
\end{figure}

\paragraph{Broadcasting strategies parameters} 
For the constant inter-transmission period policy, we make $T_p$ vary from 0 to 10 s: the trade off involves position estimation accuracy and broadcasting overhead. 
For the threshold-based policy, we consider $N_{E_{thr}}=30$ different values of the error threshold,~ranging between 0 and 42 m, until minimum position error is~achieved.

\paragraph{Scenario parameters} 
\label{par:scenario_parameters}
For our simulations, we use a real road map data imported from {OpenStreetMap} (OSM), an open-source software which combines wiki-like user generated data with publicly available information. We consider the OSM map of New York City, as represented in Fig.~\ref{fig:osm}, to characterize a very dynamic urban environment, and the total simulation area is set to $A_S=0.5168$ km$^2$.
Moreover, in order to consider realistic mobility routes that are representative of the behavior of vehicles  in the VANET, we simulate the mobility of cars using SUMO, as represented in Fig.~\ref{fig:sumo}.
The vehicles move through the street network  according to a \texttt{randomTrip} mobility model, which generates trips with random origins and destinations, and speeds which depend on the realistic interaction of the vehicle with the road and network elements. The maximum  speed is set to $v_{max}=13.89$ m/s, which is consistent with an urban scenario.
A density of $d_v=120$ vehicles/km$^2$ is considered, making $|V|=62$ the number of vehicles deployed in the considered VANET~scenario.

\paragraph{Evaluation metrics} 
\label{par:evaluation_metrics}
The performance of the proposed broadcasting strategy is assessed in terms of
\begin{itemize}
  \item \gls{ctra} model accuracy, i.e., how well the \gls{ctra} model  tracks vehicle mobility;
  \item Average positioning error, i.e., the average error committed by the ego vehicle to estimate its own state, i.e., geographical position, and that of its neighbors;
  \item 95th percentile of the positioning error, i.e., the average positioning error relative to the worst 5\%  of the vehicles;
  \item Detection error, i.e., the sum of the undetection (i.e., unknown vehicles in the neighborhood) and misdetection (i.e., vehicles that are believed to be in the area but are outside the communication range) event probabilities.\\
\end{itemize}

The remainder of this section is organized as follows: we evaluate the performance of the \gls{ctra} motion model considering long- and short-term tracking in Sec.~\ref{sub:CTRA}, while in Sec.~\ref{ssec: perf} we compare the position estimation accuracy of the proposed threshold-based broadcasting strategy against a constant inter-transmission period scheme.

\subsection{CTRA analysis}\label{sub:CTRA}
The quality of the tracking and the accuracy of the \gls{ctra} approach strongly depend on the scenario. A model such as \gls{ctra} is designed to deal with slow variations, and rapid changes in the acceleration are extremely difficult for it to~track.

Intuitively, a regular and almost time-invariant scenario seems to fit this kind of model better.
In such a scenario, a regular transmission strategy might not be too damaging, since the \gls{ctra} model is essentially correct; the errors are predictable within the model, as well as relatively slow. 
In a more dynamic scenario, such as the urban one we consider, several elements make the environment more unpredictable: traffic or pedestrians might cause the driver to brake sharply, and navigation is not simple, as turns, crossings and traffic lights make the vehicle brake, turn or accelerate suddenly. These discontinuities reduce the accuracy of the model, showing its limits in complex scenarios. Finding out their frequency and severity is then extremely important in the design of both improved tracking algorithms and more efficient information dissemination strategies: the final objective is to track vehicles in the \gls{vanet} accurately with limited signaling overhead.

The accuracy of the \gls{ctra} model is a way to determine the relation between time and \gls{qoi}: if the model is accurate, the decay of information follows a known distribution, and periodic transmissions can optimize \gls{qoi} reasonably well. Inaccuracies in the model reflect an increased randomness in the relation.

In order to compare the results of the \gls{ctra} tracking model with the empirical reality, we can use the inter-transmission interval as a proxy for the accuracy of the \gls{ctra} model: if we assume a threshold-based transmission policy, the theoretical inter-transmission interval distribution is determined by the parameters of the \gls{ukf}. The error on the vehicle position tracking after $i$ \gls{ukf} prediction steps is a multivariate Gaussian random variable with zero mean and a semidefinite positive covariance matrix $P_i$. The probability that the error $e_i$ after $i$ \gls{ukf} prediction steps  is higher than the threshold $E_{\text{thr}}$ is then given by the inverse \glspl{cdf} of the position error over the circle with radius $E_{\text{thr}}$:
\begin{align}
 P(e_i>E_{\text{thr}})&=1-\int_{B(E_{\text{thr}})}\phi\left((x,y)\right)d(x,y),
\end{align}
where $B(E_{\text{thr}})$ represents the circle of radius $E_{\text{thr}}$, i.e, $B(E_{\text{thr}})=\{(x,y):\sqrt{x^2+y^2}\leq E_{\text{thr}}\}$, and $\phi((x,y))$ is the bivariate normal \gls{pdf}. We now need to calculate the probability of going over the threshold for the first time after $i$ steps:
\begin{align}
 P_{\text{tx}}(i)&=P(e_i>E_{\text{thr}})\prod_{j=1}^{i-1}(1-P(e_j>E_{\text{thr}})).
\end{align}
This is a slight approximation, since $P(e_i>x)$ is actually different from $P(e_i>x|e_{i-1}\leq x)$; in the former case the position error is multivariate Gaussian, while in the latter the real error on each axis is the sum of a multivariate Gaussian and a truncated multivariate Gaussian variable, which is considerably more complex to compute numerically. However, the error introduced by this approximation is minimal, since the first step is the one with the lowest variance. It is also biased in a conservative direction: the approximation slightly overestimates the error compared to the actual model.

\begin{figure}[t]
  \centering
  \setlength{\belowcaptionskip}{-0.5cm}
  \includegraphics[width=2.8in]{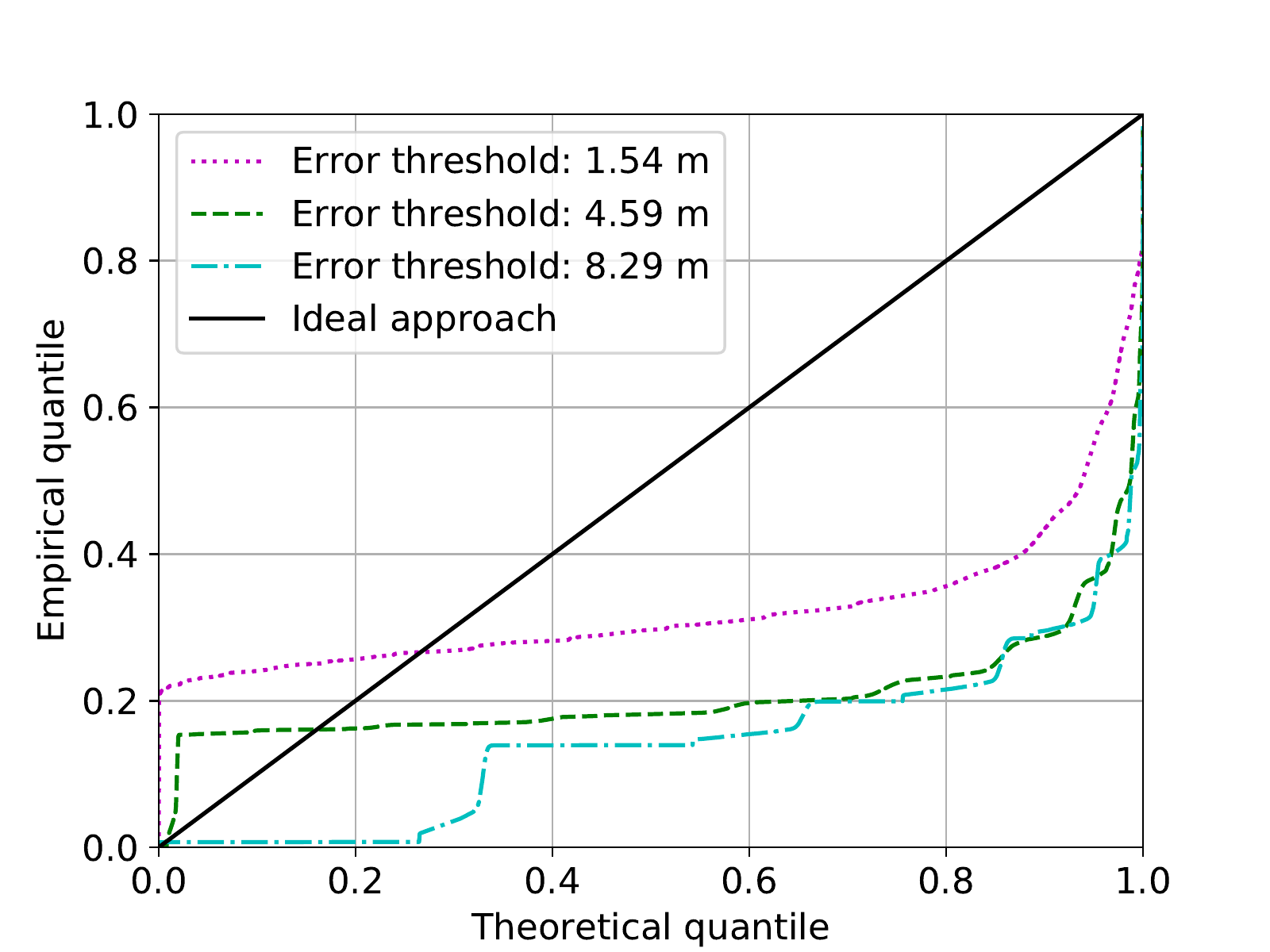}
  \caption{Q-Q plot comparing the empirical inter-transmission time distribution and the one estimated by the \gls{ctra} model for the threshold-based policy.}
  \label{fig:qqplot}
\end{figure}

The theoretical inter-transmission time distribution we defined above is compared with the empirical distribution obtained from our simulations in Fig.~\ref{fig:qqplot}, which shows a Q-Q plot of the two with three different threshold $E_{thr}$ parameters: 1.94 m, 4.59 m, and 8.29 m, sorted by increasing average inter-transmission time. 
The results are obtained in an ideal scenario without the effects of the channel conditions. 
The plot shows that the \gls{ctra} model evolves inaccurately with time, consistently overestimating the long-term error. There are two interesting effects in the plot: firstly, the time distribution calculated by \gls{ctra} does not model the actual behavior of the system (black diagonal) very well, and secondly, there is a significant number of transmissions (around 20\% for the $E_{thr}=1.94$ m and $E_{thr}=4.59$ m cases) performed almost immediately. 
The former highlights the well-known issues of the \gls{ctra} model in long-term prediction~\cite{xie2018vehicle}, while the latter is probably caused by unpredictable maneuvers such as lane switches, turns or hard brakes. This explanation is supported by the absence of immediate transmissions for the system with the 8.29 m threshold, which is probably wide enough to avoid triggering a transmission after one of these maneuvers.

We conclude that a periodic transmission policy based on the \gls{ctra} parameters would consistently overestimate the system error and consequently underperform, while a threshold-based approach that considers the tracking error directly would simply transmit whenever the error is above the threshold, increasing the average inter-transmission time to avoid congestion and compensating for the systematic error. 

\subsection{Performance comparison}\label{ssec: perf}
\begin{figure}[t!]
\centering
 \begin{subfigure}[t]{0.93\columnwidth}
  \centering
  \includegraphics[width=0.85\columnwidth]{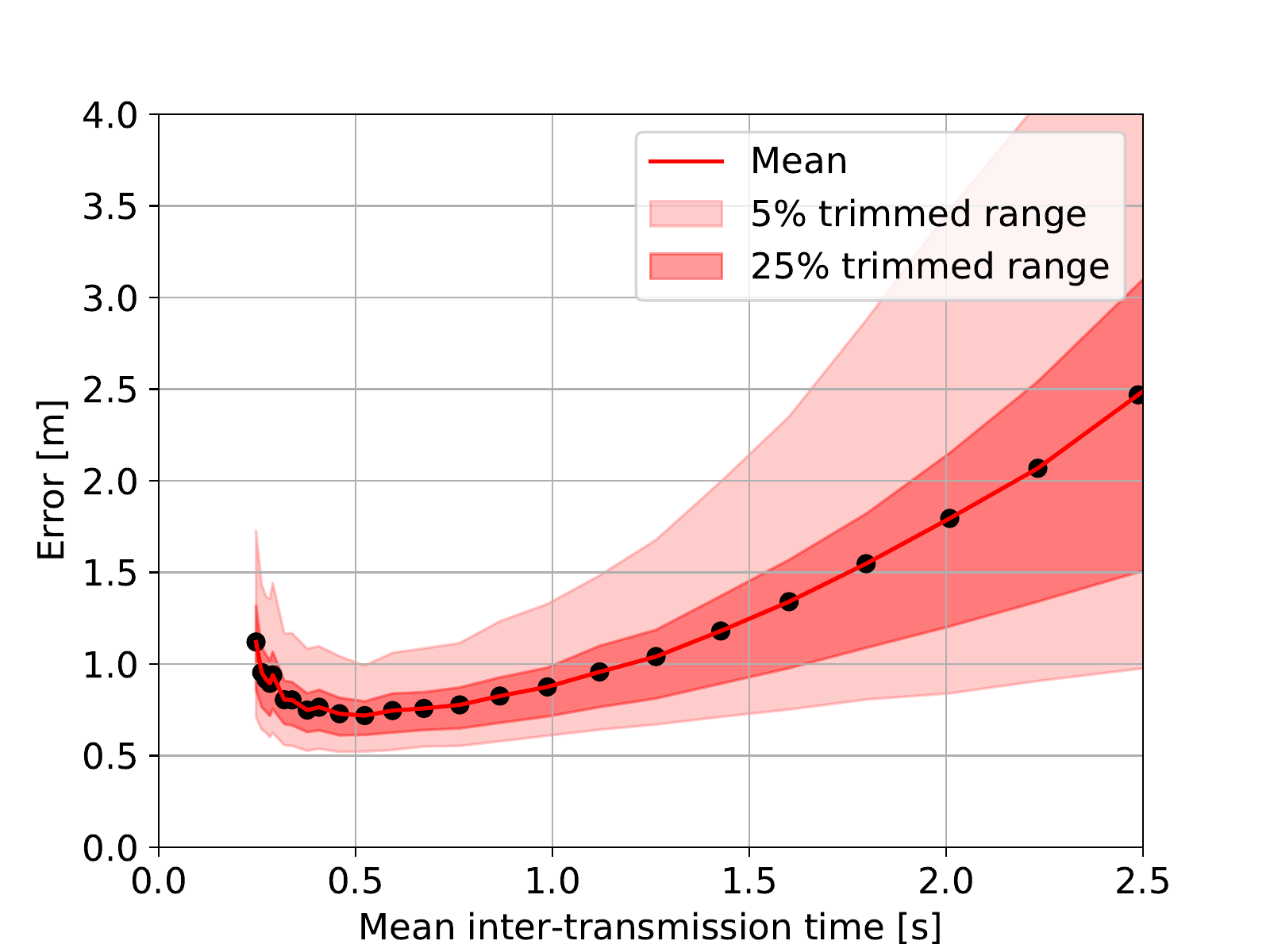}
    \caption{Constant inter-transmission period  policy.}
  \label{fig:pol0}
  \end{subfigure} 
  \begin{subfigure}[t]{0.93\columnwidth}
  \centering
  \includegraphics[width=0.85\columnwidth]{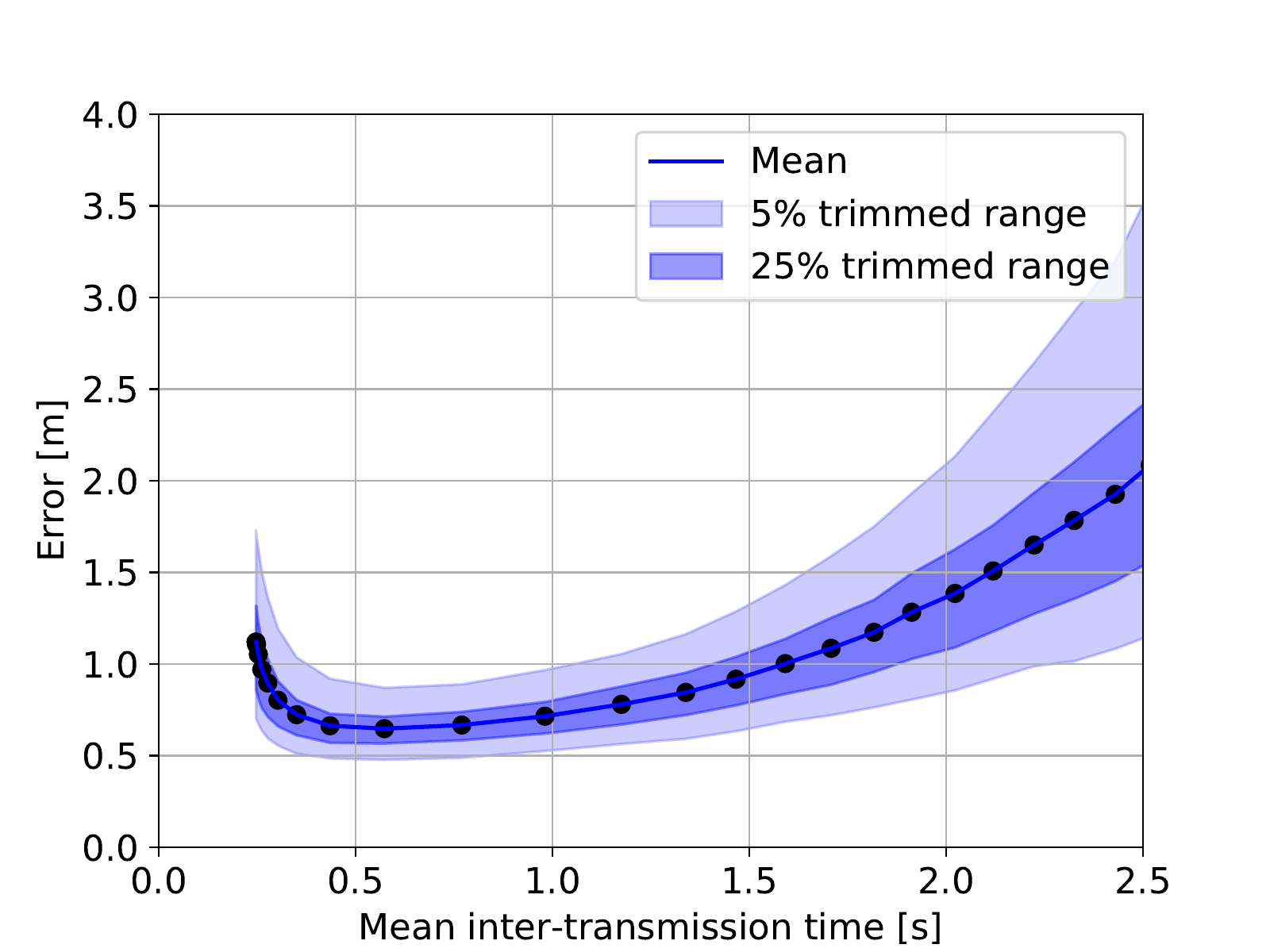}
  \label{fig:pol1}
  \caption{Threshold-based policy.}
  \end{subfigure}
  \setlength{\belowcaptionskip}{-0.5cm}
  \caption{Comparison of the error distributions as a function of the average inter-transmission time and the information broadcasting strategy.}
  \label{fig:policies}
  \vspace{-0.1cm}
\end{figure}

\begin{figure}[t!]
\centering
 \begin{subfigure}[t]{0.49\columnwidth}
  \centering
  \includegraphics[width=0.99\columnwidth]{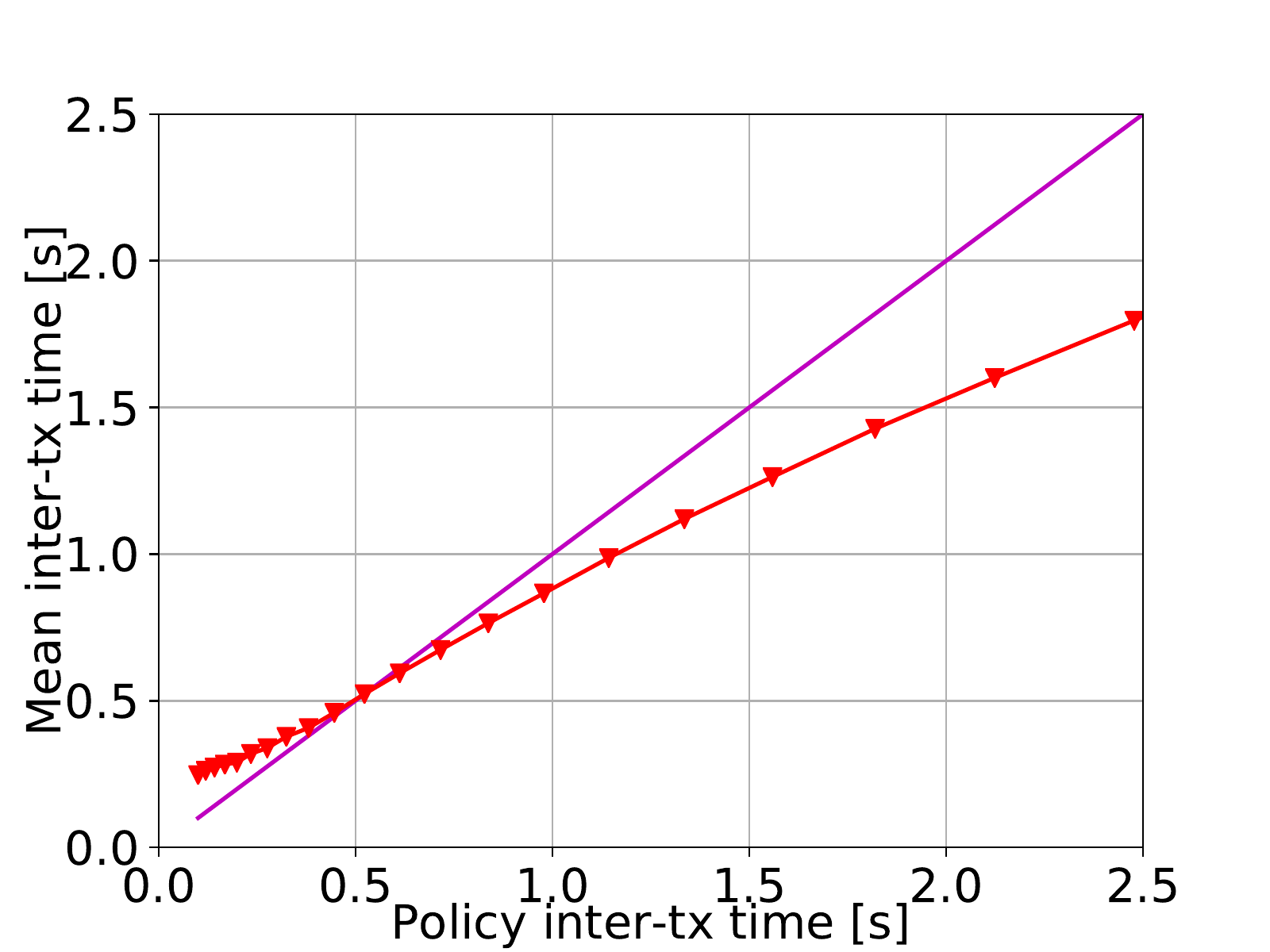}
      \caption{Constant inter-transmission period.}
  \label{fig:intertx_p0}
  \end{subfigure} 
  \begin{subfigure}[t]{0.49\columnwidth}
  \centering
  \includegraphics[width=0.99\columnwidth]{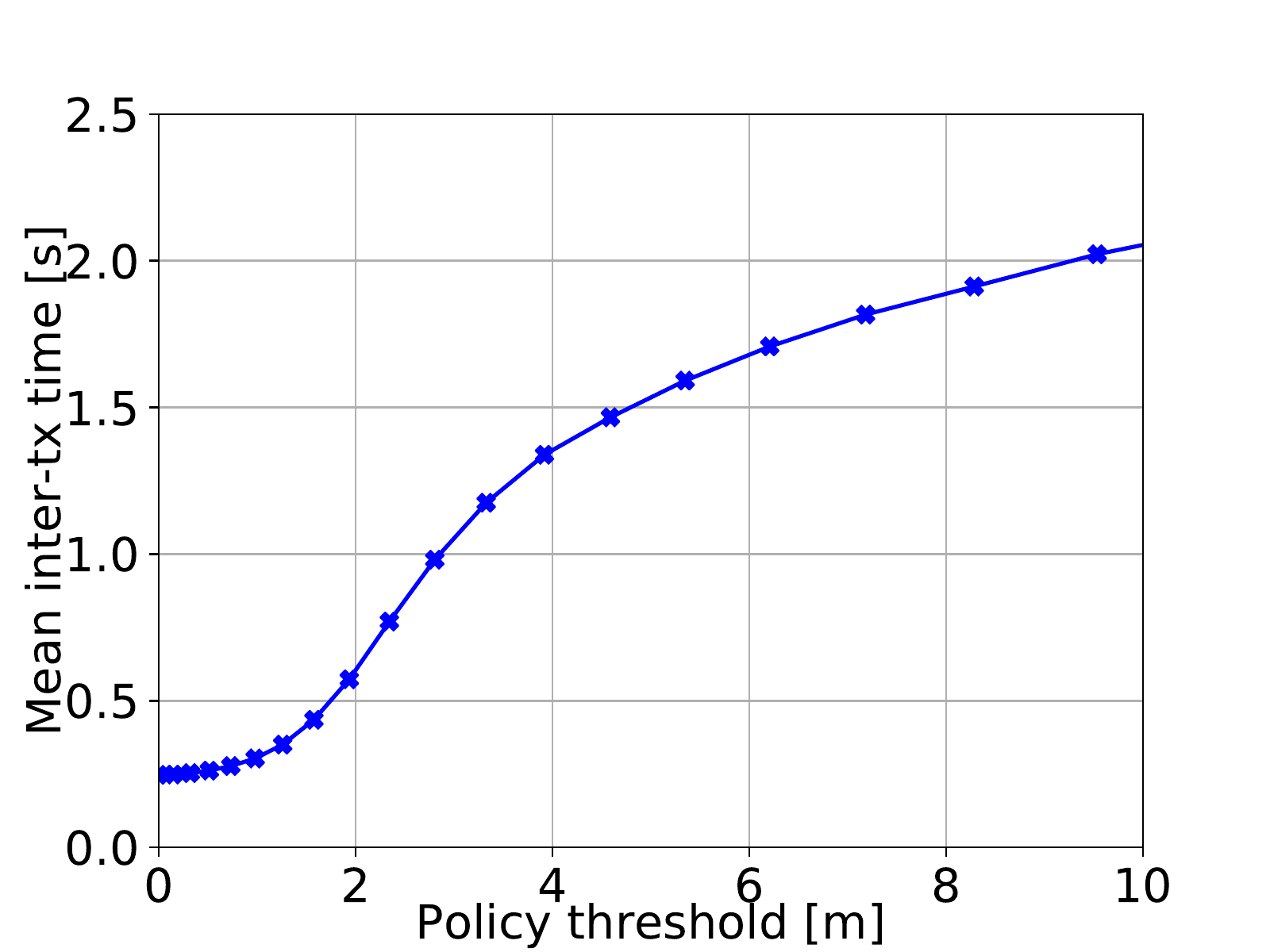}
  \caption{Threshold-based.}   
  \label{fig:intertx_p1}
  \end{subfigure}
  \setlength{\belowcaptionskip}{-0.5cm}
      \caption{Average inter-transmission time as a function of policy parameters.}
  \label{fig:intertx}
\end{figure}

The performance of a policy is not easy to compare: since the positioning error for any policy is a distribution, there is no single parameter that defines which one is better.
However, a comparison of the plots in Fig.~\ref{fig:policies} shows that the threshold-based policy performs better than one considering constant inter-transmission  periods: both the mean and the variance of the error are lower, resulting in a more accurate picture of the state of the \gls{vanet}. 

We highlight that the mean inter-transmission time shown on the $x$-axis of Fig.~\ref{fig:policies} and all subsequent ones takes medium access effects into consideration. 
As stated in Sec.~\ref{sub:broadcasting_strategies}, the two policies can indeed either transmit more often than set to react to the appearance of new vehicles in the communication range, or their broadcasts can be delayed by the medium access mechanism. Fig.~\ref{fig:intertx} shows the actual average inter-transmission time as a function of the policy parameter. 
Even the periodic policy does not strictly respect its set period, shown as the purple diagonal in Fig.~\ref{fig:intertx_p0}: it generally transmits more often because of the unscheduled transmissions when new vehicles are detected, but reducing the policy's inter-transmission period below a certain point leads to delayed transmissions due
to the limits of \gls{csmaca} (e.g., vehicles that try to transmit every 100 ms will have an effective inter-transmission time of about 250 ms).\footnote{Besides \gls{csmaca}, other ad hoc contention methods can be implemented to decrease the collision probability and guarantee an improved packet delivery performance. While of interest in its own right, such analysis is beyond the scope of this paper and is left for future work.}
The same pattern can be seen for the threshold-based policy in Fig.~\ref{fig:intertx_p1}: reducing the error threshold to values below 1 m has no effect on the inter-transmission time, because updates are then delayed because of congestion.
These patterns are also evident in the trends of the policy error in Fig.~\ref{fig:policies}: the curves for both policies are convex, exemplifying the existence of an optimal transmission level which guarantees minimum positioning error within the~network. Attempting to access the channel more often will increase the overall error, since delayed transmissions and collisions will become the main limiting factor to the system accuracy.

\begin{figure}[t]
  \centering
  \setlength{\belowcaptionskip}{-0.5cm}
  \includegraphics[width=2.95in]{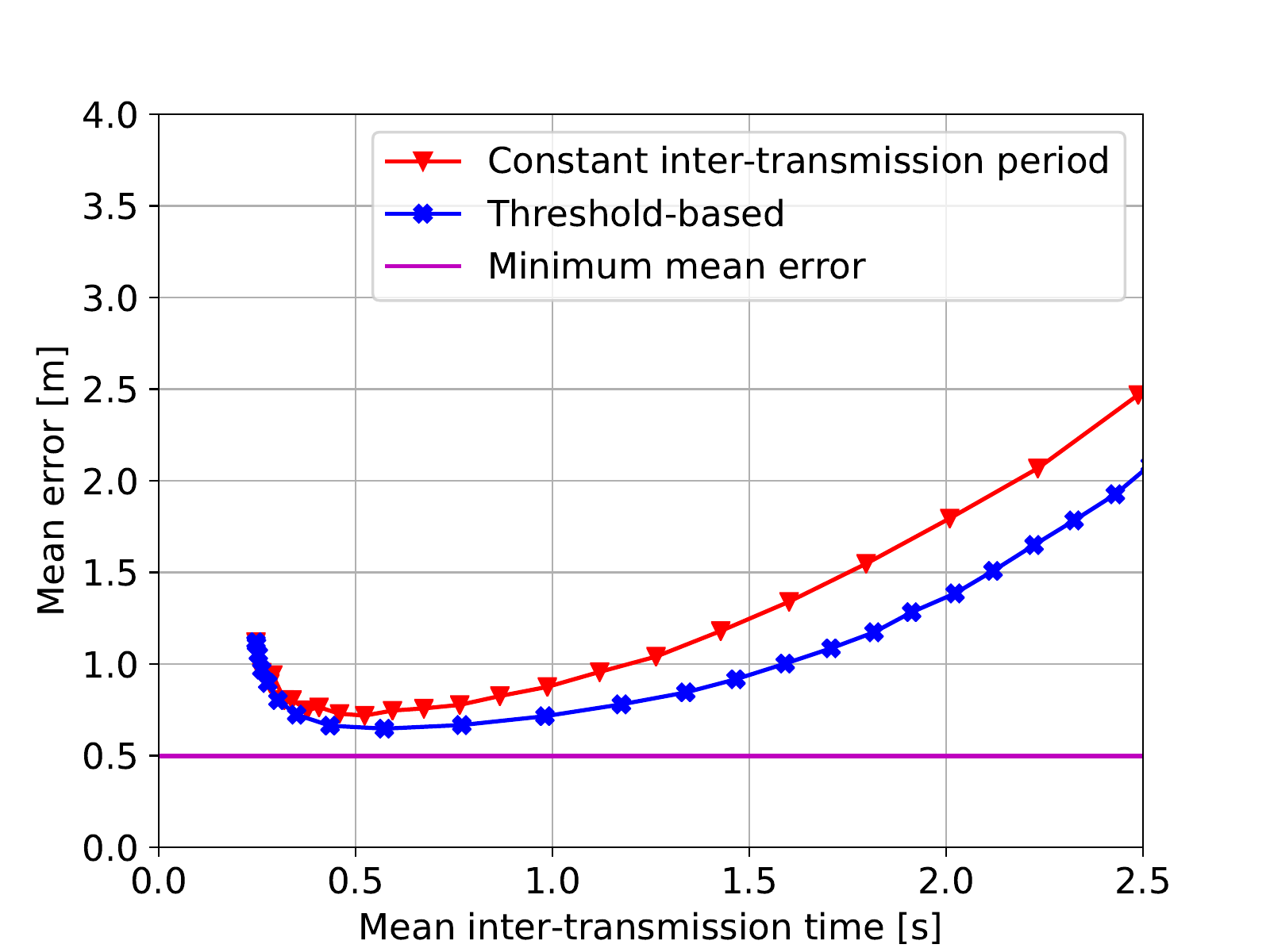}
  \caption{Average positioning error as a function of the average inter-transmission time and the information broadcasting strategy.}
  \label{fig:error_mean}
\end{figure}

\begin{figure}[t]
  \centering
  \setlength{\belowcaptionskip}{-0.5cm}
  \includegraphics[width=2.95in]{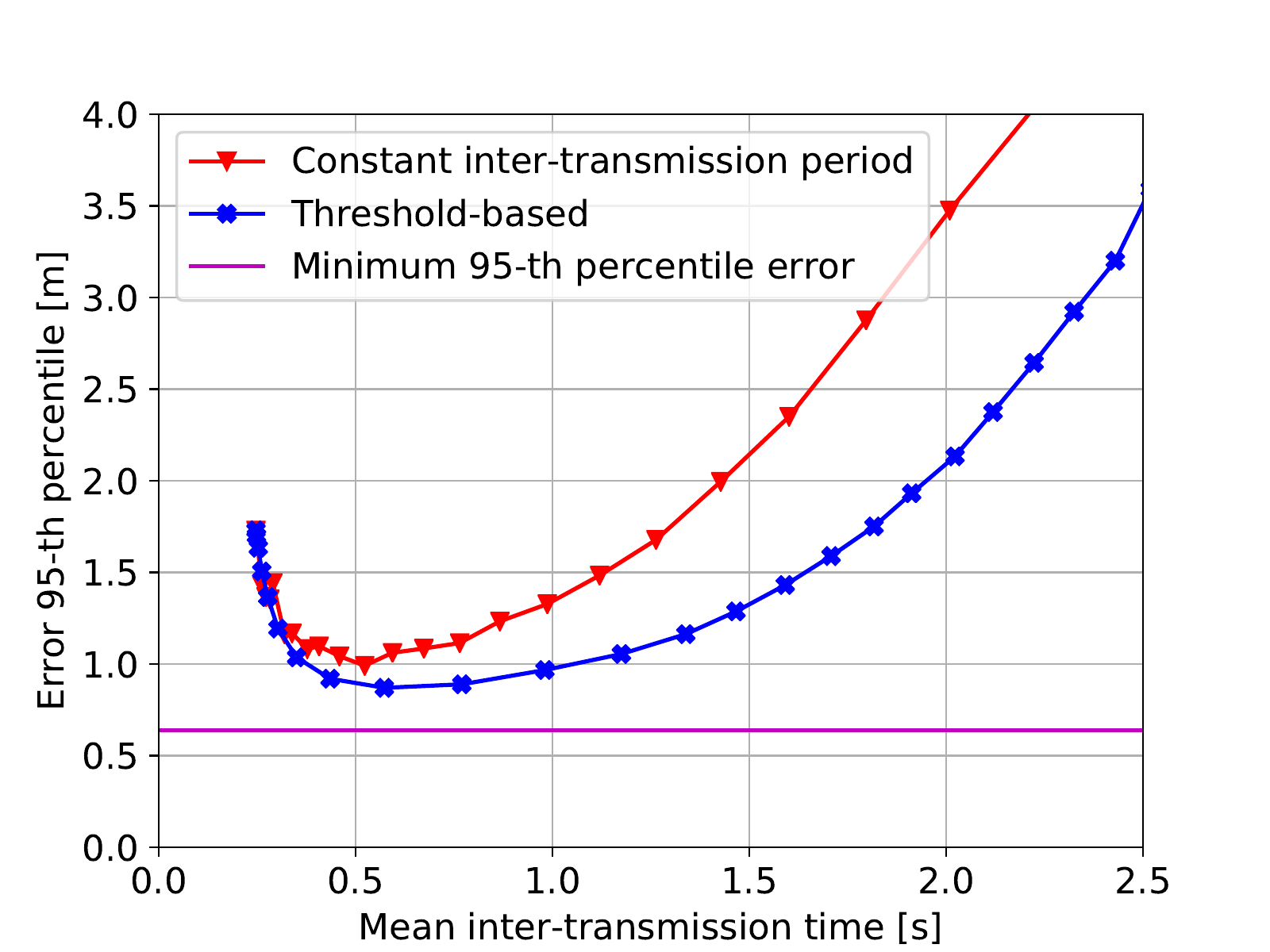}
  \caption{95th percentile of the positioning error as a function of the average inter-transmission time and the information broadcasting strategy.}
  \label{fig:error_95}
\end{figure}

Fig.~\ref{fig:error_mean} shows the average error for the two policies, compared to the minimum ideal bound, which is the average error in a \gls{vanet} in which all vehicles broadcast their positioning information every 100 ms, i.e., every timeslot, on an ideal channel with no packet losses or delays due to the medium access. 
As the figure shows, the average error achieved by the threshold-based policy is always lower than that of the periodic transmission strategy, significantly reducing the optimality gap by using the channel only when necessary. Furthermore, the performance gap is even more significant when the threshold is increased, reducing the transmission frequency. In a congested \gls{vanet} scenario in which multiple applications need access to vehicular communications, reducing network load without compromising positioning accuracy would significantly improve the system performance. The same considerations are valid when we use a worst-case approach: Fig.~\ref{fig:error_95} shows the 95th percentile of the position estimation error as a function of the average inter-transmission time, and the patterns we observed in the previous plot are even more evident.
The above discussion motivates efforts towards the design of \gls{qoi}-aware mechanisms that take quality of information, rather than age, as the decision factor for broadcasting. 

\begin{figure}[t]
  \centering
  \setlength{\belowcaptionskip}{-0.5cm}
  \includegraphics[width=2.95in]{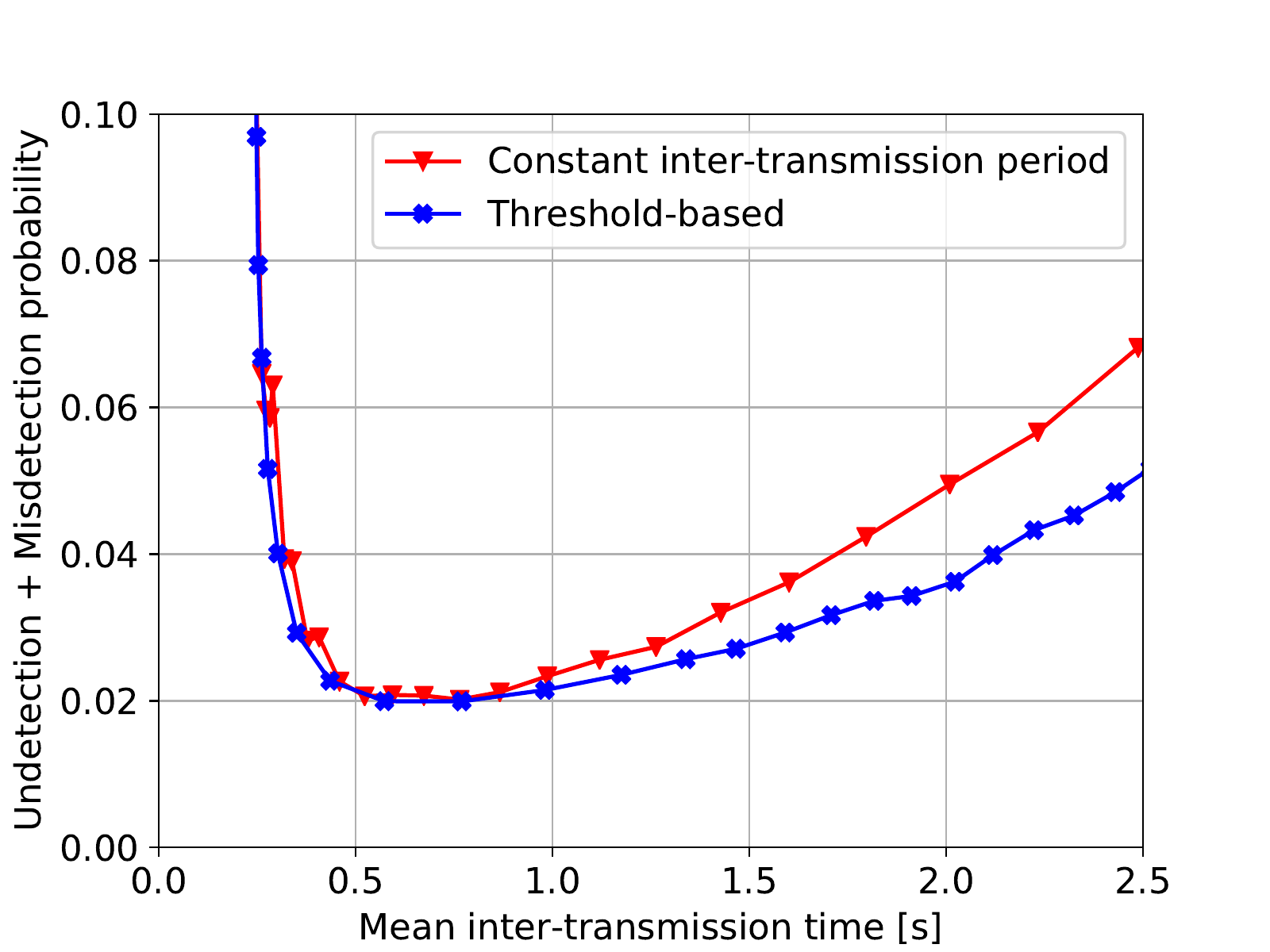}
  \caption{Detection error as a function of the average inter-transmission time and the information broadcasting strategy.}
  \label{fig:detect}
\end{figure}

Finally, the probability of having undetection (i.e., unknown vehicles in the neighborhood) or misdetection events (i.e., vehicles that are believed to be in the area but are outside the communication range) is another important system parameter. If a vehicle is wrong about its neighbors, its local dynamic map will be wrong in unpredictable ways. Fig.~\ref{fig:detect} shows the sum of these two probabilities, measured as the number of events over the number of neighbors. The policies have similar detection error probabilities, which are mostly due to undetection events: the probability increases with the inter-transmission time, since transmissions are sparser and vehicles can enter the neighborhood without being detected, but it also becomes very high when vehicles transmit very often: the packet losses and transmission delays due to \gls{mac} congestion can effectively hide vehicles, causing a significant undetection~problem.

\section{Conclusions and future work}

In this work, we have investigated the possibility of improving position estimation and prediction in \glspl{vanet} by implementing a threshold-based broadcasting strategy. Nowadays, most  \gls{vanet} applications exploit tracking systems in which the target state coincides with the set of the vehicle positions. To avoid the decay of information, such systems need to be very frequently updated with the data collected through inter-vehicular communications. This conflicts with the specific characteristics of the \gls{vanet} scenario, in which the channel access cannot be centrally managed and transmissions are affected by packet collisions. 

To study this trade-off, we built a comprehensive \gls{vanet} model, which uses \gls{sumo} to incorporate realistic vehicle mobility traces in an urban scenario. 
Our results show that tracking systems based on the \gls{ctra} motion model ensure good forecasting performance only for short-time predictions. 
We also demonstrated through simulation that the proposed threshold-based strategy, which considers \gls{qoi} as the decision factor for disseminating positioning information, outperforms a benchmark scheme which broadcasts positioning updates periodically, since it reduces the communication overhead while ensuring the same forecasting~performance.  

As part of our future work, we will analyze the relation between the optimal system performance and the broadcast strategy settings, so that vehicles can autonomously adapt their behavior according to the network dynamics. We will also improve our model framework by implementing congestion control mechanisms. Finally, we are interested in developing more advanced communication strategies, e.g., by exploiting \gls{ml}, targeting improved tracking~accuracy. 

\bibliography{bibliography}
\bibliographystyle{IEEEtran}
\end{document}